\documentclass[useAMS,usenatbib,preprint2,longauthor]{aastex61}
\usepackage{natbib}
\usepackage{amsmath}
\usepackage{amssymb}
\usepackage{graphicx}
\usepackage{color}
\usepackage{multirow}
\usepackage{enumerate}
\usepackage{float}
\usepackage{comment}
\usepackage{enumerate}
\usepackage{placeins}
\usepackage{subfigure}
\begin{document}
\title[Accretion disc around PSR B1259-63]{A new approach to the G\MakeLowercase{e}V flare of PSR B1259-63/LS2883}
\author{Shu-Xu Yi$^\dagger$}
\affiliation{Department of Physics, The University of Hong Kong, Pokfulam Road, Hong Kong}
\author{K. S. Cheng$^\star$}
\affiliation{Department of Physics, The University of Hong Kong, Pokfulam Road, Hong Kong}
\email{$\dagger$:yishuxu@hku.hk}
\email{$\star$:hrspksc@hku.hk}

\begin{abstract}
PSR B1259-63/LS2883 is a binary system composed of a pulsar and a Be star. The Be star has an equatorial circumstellar disk (CD). The {\it Fermi} satellite discovered unexpected gamma-ray flares around 30 days after the last two periastron passages. The origin of the flares remain puzzling. In this work, we explore the possibility that, the GeV flares are consequences of inverse Compton-scattering of soft photons by the pulsar wind. The soft photons are from an accretion disk around the pulsar, which is composed by the matter from CD captured by the pulsar's gravity at disk-crossing before the periastron. At the other disk-crossing after the periastron, the density of the CD is not high enough so that accretion is prevented by the pulsar wind shock. This model can reproduce the observed SEDs and light curves satisfactorily.
\end{abstract}

\section{Introduction}
PSR B1259-63/LS 2883 is a binary system composed of a radio pulsar and a massive main sequence Be star \citep{1992ApJ...387L..37J}. The pulsar PSR B1259-63 is a non-recycled, spin-down powered radio pulsar, with a spin period of 47.76\,ms, and a period derivative of 2.27$\times10^{-15}$ \citep{2014MNRAS.437.3255S}. The companion star is a Be star with the mass of $\sim31\,M_\odot$, and rotating at a near-break-up rate \citep{2011ApJ...732L..11N}. The binary orbit is highly eccentric, with the eccentricity $e=0.87$, orbital period $P=1237$\,days and semimajor axis $a=7.2$\,AU. The stellar wind of the Be star interacts with the pulsar wind fiercely, forming a termination shock front, which is believed to be the origin of un-pulsed, orbital modulated high energy radiations of X-rays \citep{1999ApJ...521..718H,2006MNRAS.367.1201C} and TeV gamma-rays \citep{1999APh....10...31K,2005A&A...442....1A}. The ejected matter at the equator of the Be star builds up a Keplerian decretion disk, which transports the angular momentum of the star outwardly \citep{2013A&ARv..21...69R}. The obstruction of the circumstellar disk (CD) leads to eclipse of the pulse radio emission near the periastron \citep{1992ApJ...387L..37J,1996MNRAS.279.1026J}, and the interaction between the pulsar and the CD gives rise to the double peaks structure of the X-ray light curve \citep{2011MNRAS.416.1067K}.

During the periastron passage in 2010/2011, the first passage after the launch of the {\it Fermi} satellite, \cite{2011ApJ...736L..10T} and \cite{2011ApJ...736L..11A} discovered the gamma-rays emission (100\,MeV-100\,GeV) from the system with the {\it Fermi}-LAT detector. Marginal gamma-rays were detected within period around the time of the periastron $t_{\rm{p}}$, before it faded at $\sim t_{\rm{p}}+15$ days. Then at $\sim t_{\rm{p}}+30$ days, an sudden re-emergence of the gamma-ray emission surprisingly occurred, with the flux fast rising up to $20-30$ times of its previous value over a few days, followed by a slow decay about two weeks before its disappear after $~t_{\rm{p}}+60$ days. The GeV flare reappeared during the following periastron passage in 2014 \citep{2014ATel.6204....1M,2014ATel.6216....1T,2014ATel.6225....1W} at approximate the same orbital phase. The similarities of the flux, on-set phases, duration and sub-scale structures in the light curves between GeV flares in two periastron passages \citep{2015ApJ...798L..26T,2015ApJ...811...68C} suggest its repeating nature and the origin due to the interaction between the pulsar and the circumstellar matter. The cause of the GeV flare is puzzling: as the first-considered radiation mechanism, inverse Compton scattering (IC) of the cool pulsar wind electrons with seed photons from either the Be star or the CD gives the maximum of GeV flux at the periastron, where the external photons field is densest\footnote{Considered the anisotropy of IC, the GeV light curve peaks slightly before the periastron.}. Whereas the flare occurs $\sim30$\,days after the periastron \citep{2007MNRAS.380..320K,2009ApJ...702..100T,2011MNRAS.417..532P,2012MNRAS.426.3135V}
; on the other hand, if we trust the geometry of the CD implied by the X-ray light curve \citep{2006MNRAS.367.1201C}, then the flare appears when the pulsar has well left the CD. Therefore it is difficult to attribute the flare to the immediate interaction between the pulsar and the CD.

The mystery of the GeV flare attracts many explaining attempts: \cite{2012ApJ...752L..17K} still tried to explain the flare as a consequence of cool pulsar wind IC with the soft photons from the CD, but they also took into account of the time variance of the pulsar wind zone length (PWZ) towards the observer. PWZ is the place where the IC process take place. They argued that the rapid rising of the gamma-ray flux corresponds to the fast growth of the PWZ immediately after the pulsar's exiting from the CD. The main difficulty of this model is insufficient soft photons from CD for IC. As mentioned by the authors, the required target photon luminosity is $\sim40\%$ of LS 2883. Significant heating of the CD by the passage of the pulsar is supposed by the authors to reconcile this problem. 

Using also IC, \cite{2013A&A...557A.127D} sought for seed photons in X-rays from the synchrotron radiation of the shock-heated electrons. Since in this model the seed photons have higher energy of a few keV, the bulk Lorentz factor of the pulsar wind should be $\sim500$ to give the observed gamma-ray emission. The Lorentz factor is much less than the typical values of pre-shock pulsar wind (see \citealt{2012MNRAS.426.3135V} and references therein). It is difficult for this model to explain the delay between the GeV flare and the X-ray lightcurve maximum.

Another approach to the GeV flare, proposed by \cite{2011ApJ...736L..10T} and modeled in detail by \cite{2012ApJ...753..127K}, is the Doopler-boosted synchrotron radiation in the shock tail: The stellar wind from the Be star collides with the pulsar wind to form a termination shock. Electrons (and positrons) in the pulsar wind are shock-heated and flow along the shock-tail at a mildly relativistic velocity. The hot electrons radiate away their energy through the synchrotron radiation, which is in X-ray in the flow co-moving frame and is Doopler-boosted to gamma-ray at the rest frame. This model is favored by the observational fact that the GeV flare appears when the pulsar is around the inferior conjunction, where the shock tail is close to the line of sight. However this model predicts that the light curves of the X-ray and gamma-ray peak at the same orbital phase, which violates the observation.

\cite{2013ApJ...776...40M} explored the possibility that, the flare is the result of a transit between the driven reconnection and the electromagnetic precursor scenarios of the termination shock. The above mentioned scenarios are both ways in which the oscillating magnetic field of the rotating pulsar dissipates its energy. The radiation efficiency is low in the driven reconnection scenario, and is high in the electromagnetic precursor scenario. When the condition of transition is satisfied, the radiation efficiency jumps thus gives the observed flare. This model also predicts another GeV flare at a different orbital phase when the transition condition is satisfied again (the peak is not necessary in the GeV band as mentioned by the authors, depending on the wind conditions.) The prediction is not supported yet by observations.

A recent work of \cite{2017A&A...598A..13D} studied numerically the impact of stellar wind clumps or inhomogeneities on the high-energy non-thermal radiation in binary system like this. They applied their study
to PSR B1259-63/LS2883, and explored the scenario that a dense matter clump impacts on the two-wind interaction region. Although enlightening, their simulation is not able to reproduce the GeV flare satisfactorily.

We explore the possibility in this paper that, when the pulsar passes across the CD, some of the matter in the CD is captured by the gravity of the neutron star. After the pulsar exits the CD, the captured matter forms an accretion disk. Soft photons emitted from the accretion disk are up-scattered by the pulsar wind, results in the observed gamma-ray flares. The gravity capture happens only when the termination shock front are inside the capture radius, inside which the velocity of the CD matter in the pulsar rest frame is less than the escape velocity from the gravity field of the pulsar. If the radius of the termination shock front is larger than the capture radius, the matter flow will be redirected before it enters the capture radius. We should also check whether the captured matter has enough specific angular momenta to form an accretion disk. The detailed calculations in section 2.1 show that the formation of an accretion disk by this scenario is likely.

This paper is organized as follows. In section 2 we describe the model: In section 2.1 we study how the CD matter is captured by the gravity of the pulsar; Then in section 2.2 we describe how an accretion disk is formed from the captured matter, and how the accretion disk evolves with time; In section 2.3 we study the IC process, in which the pulsar wind up-scatters the accretion disk soft photons to gamma-rays. In section 3 we compare our model calculations with observations. We conclude and discuss in section 4.

\section{Model}
\subsection{Accretion from the stellar disc}
To describe the motion of the pulsar and the CD matter, we construct the Cartesian coordinates frame as follows: the origin is on the barycenter of Be star and the pulsar, the $x$-axis is towards the periastron, $z$-axis is along the angular momentum of the orbit. The norm vector of the Be star's CD is $$\mathbf{n_{\rm{cd}}}=(\sin\theta_{\rm{n}}\cos\phi_{\rm{n}},\sin\theta_{\rm{n}}\sin\phi_{\rm{n}},\cos\theta_{\rm{n}}).$$ Where $\theta_{\rm{n}}$ and $\phi_{\rm{n}}$ are the polar angle and the azimuthal angle of the norm vector respectively. The pulsar intersects with the mid-plane of the CD at the true anomaly $\phi_{\rm{cd,\pm}}=\phi_n\pm\pi/2$. Based on the Kepler equations, the velocity of the pulsar as a function of the true anomaly $\phi$ is:
\begin{equation}
\begin{array}{l}
v_{\rm{p},x}=e\sqrt{p\mu}\sin\phi\cos\phi/p-\sin\phi\sqrt{p\mu}/r\\
v_{\rm{p},y}=e\sqrt{p\mu}\sin^2\phi/p+\cos\phi\sqrt{p\mu}/r\\
\label{eqn:first}
\end{array}
\end{equation}
where $r=p/(1+e\cos\phi)$ is the pulsar-star distance, $e$ is the eccentricity of the orbit, $\mu\equiv G(M_\star+M_{\rm{p}})$ is the gravity constant times the sum of the total mass of the Be star and the pulsar, $p\equiv a(1-e^2)$ and $a$ is the semi-major axis.
The CD is thought to be Keplerian, therefore we assume the velocity of the disk matter is pure tangential and ignore the outward velocity. As a result, at any point $\mathbf{r}$ the velocity of the disk matter is:
\begin{equation}
\mathbf{v_{\rm{cd}}}=\sqrt{\frac{GM_\star}{r}}\mathbf{n}_{\rm{cd}}\times\mathbf{r}/r.
\label{eqn:vsd}
\end{equation}
Here we presume that the rotation of CD is prograde with the orbit of the pulsar. The retrograde case is discussed later.
The velocity of the disk matter viewed at the pulsar rest frame is $\mathbf{v_{\rm{rel}}}=\mathbf{v_{\rm{cd}}}-\mathbf{v}_{\rm{p}}$. Any matter with the impact parameter less than the radius
\begin{equation}
r_{\rm{BH}}=\frac{2GM_{\rm{p}}}{v^2_{\rm{rel}}},
\label{eqn:rc}
\end{equation}
is thought to be captured by the gravity of the neutron star. $r_{\rm{BH}}$ is the Bondi-Hoyle value \citep{1944MNRAS.104..273B} when $v_{\rm{rel}}$ is much larger than the sound speed. Mass transfer via gravity-capture is only possible when the shock front is within $r_{\rm{BH}}$, i.e., $r_{\rm{s}}<r_{\rm{BH}}$ as discussed in above section, where
\begin{equation}
r_{\rm{s}}=\sqrt{\frac{L_{\rm{spin}}}{4\pi\rho_{\rm{cd}}v^2_{\rm{rel}}c}},
\label{eqn:rs}
\end{equation}
and $L_{\rm{spin}}$ is the spin down power of the pulsar, $\rho_{\rm{cd}}$ is the local density of the CD.
We adopt the following parameters:
\begin{itemize}
\item{$e=0.87$, $a=7.2\,AU$}
\item{$M_\star=31\,M_\odot$ \citep{2011ApJ...732L..11N}, $M_{\rm{p}}=1.4\,M_\odot$}
\item{$\theta_{\rm{cd}}=45^\circ$, $\phi_{\rm{cd}}=19^\circ$ \citep{2011PASJ...63..893O}}
\item{$L_{\rm{spin}}=8\times10^{35}$\,ergs/s,}
\end{itemize}
The density distribution of the decretion disk of the Be stellar has been modeled by previous researchers \citep{Lee1991,Porter1999,Carciofi2006} as:
\begin{eqnarray}
\rho_{\rm{cd}}&=&\rho_0\big(\frac{R_\star}{R}\big)^n\exp\big(\frac{-z^2}{2H^2}\big)\\ \nonumber
&=&\rho_0\big(\frac{R_\star}{R}\big)^n\exp\big(\frac{-(\phi-\phi_{\rm{cd}})^2}{2\Delta\phi^2}\big),
\label{eqn:rho}
\end{eqnarray}
where $n$ is in the range $3\sim3.5$, with 3.5 corresponds to an steady state isothermal outflow, $H$ is the scale height of the disk, $z$ is the cylindrical coordinate in the vertical direction, $\Delta\phi$ is the half-opening angle of the disk projected on the orbital plane, $R_\star$ is the radius of the Be star, which is $\sim10\,R_\odot$ \citep{2011ApJ...732L..11N}. The second part of above equation is the disk density profile projected to the orbital plane, and $\Delta\phi=18.5^\circ$ as modeled by \cite{2006MNRAS.367.1201C}. 
\cite{Takata2012} argued that a large base density ($\rho_0\sim1\times10^{-9}$\,g/cm$^3$) is needed for this system to account for the double peak structure of the X-ray light curve.

\begin{figure}
\centering
\subfigure{
\includegraphics[width=8cm]{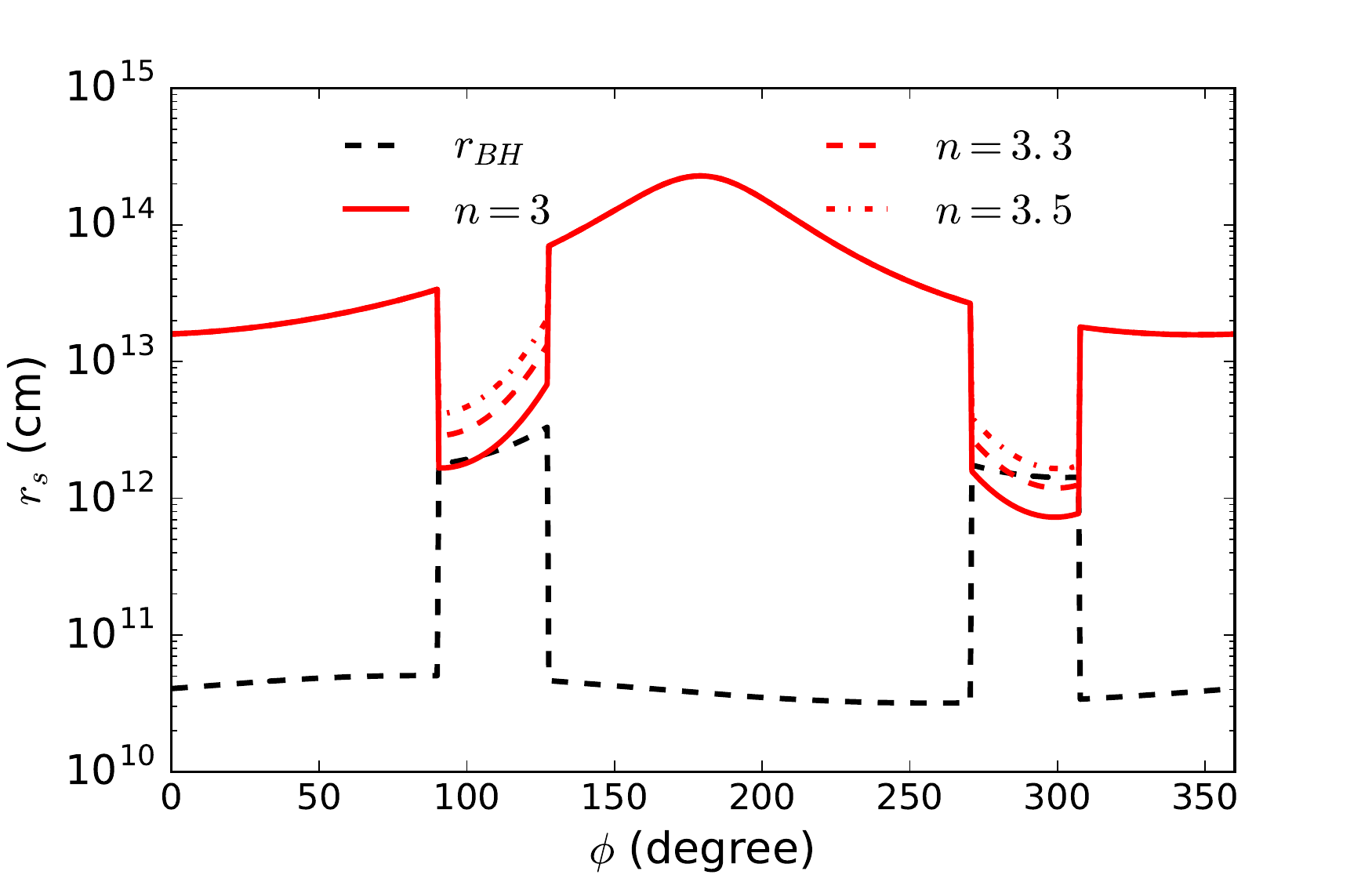}}
\subfigure{
{\includegraphics[width=8cm]{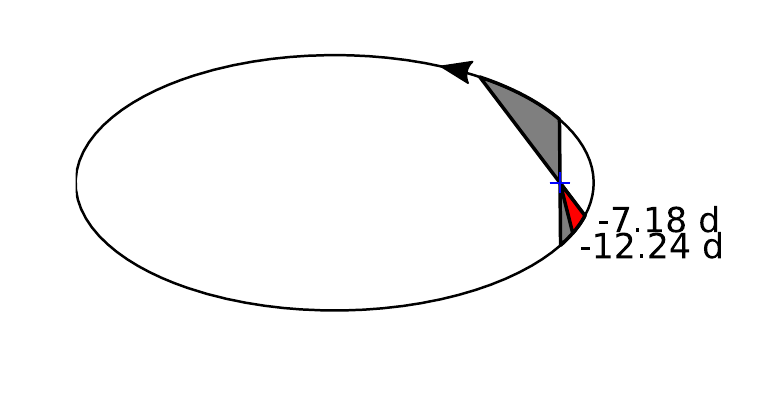}}
}
\caption{\textbf{Upper panel:} The $r_{\rm{s}}$ and $r_{\rm{BH}}$ as function of $\phi$. The red curves are $r_{\rm{s}}$ corresponding to different $n$, and the dashed black curve is $r_{\rm{BH}}$. $\phi$ is the true anomaly and the zero point is at the periastron. In both panels $\rho_0=1\times10^{-9}$\,g/cm$^3$. \textbf{Bottom panel:} The illustration of the orbit and the position of the disk (grey shade region). The arrow indicates the sense of motion of the pulsar. The red shade is the region where mass transfer takes place when $n=3.3$. The number beside the orbit is the days after periastron of the start and end of the mass transfer.}\label{fig:ell}
\end{figure}

In the upper panel of figure \ref{fig:ell}, we plot the $r_{\rm{s}}$ and $r_{\rm{BH}}$ as function of $\phi$ under different $\rho_{\rm{cd}}$ profile index $n$, with $\rho_0=10^{-9}$\,g/cm$^3$. $\phi$ is the true anomaly and the zero point is at the periastron. We ignore the perturbation of the density profile of the disk by the pulsar. We can see that the scenario of mass transfer is sensitive to $n$: when $n=3$, mass transfer occurs in both CD crossing; When $n=3.3$, mass transfer only happens during a part of the CD crossing before the periastron (``--" crossing hereafter, the other CD crossing is denoted as ``+"); When $n=3.5$, no mass transfer from the CD during both crossings. Changing the $n$ from 3 to 3.5 is equivalent to changing $\rho_0$ from $\sim$0.5 to $\sim$2 times the current value.
The illustration of the orbit and the position of the disk (grey shade region) is plotted in the bottom panel of figure \ref{fig:ell} with $n=3.3$, with the mass transfer region shaded in red.

When the condition that $r_{\rm{s}}<r_{\rm{BH}}$ is met, the Bondi-Hoyle like process transfers matter into the gravity-bounded sphere at a rate:
\begin{equation}
\dot{M}_{\rm{trans, BH}}=\pi r^2_{\rm{BH}}\rho_{\rm{cd}}v_{\rm{rel}}.
\label{eqn:mdot0}
\end{equation}
Since $v_{\rm{rel}}$ and $\rho_{\rm{cd}}$ are not uniform and have gradient across the radius of the CD, the gravity-captured matter possess the specific angular momentum of \citep{1975A&A....39..185I,1976ApJ...204..555S}:
\begin{equation}
l(t)=\frac{(GM_{\rm{p}})^2}{v^3_{\rm{rel}}}\big(\frac{|\nabla v_{\rm{rel}}|}{v_{\rm{rel}}}+\frac{|\nabla\rho_{\rm{cd}}|}{\rho_{\rm{cd}}}\big).
\label{eqn:specl}
\end{equation}

During the mass transfer process, the total accreted mass is:
\begin{equation}
M_{\rm{tot}}=\eta\int\dot{M}_{\rm{trans,BH}}dt,
\end{equation}
where $\eta$ accounts for the inefficiency of the Bondi-Hoyle accretion. According to numerical study by \cite{2013ApJ...767..135B}, under the influence of the bow shock of the pulsar $\eta$ can be as small as $\sim1\%$. Here we leave $\eta$ to be determined by fitting to the observation. \\
After ramming into the capture radius, the gases dissipate their energy and shift to the orbit of lowest energy for a given angular momentum, i.e., a torus of gas at the so called circular radius:
$R_{\rm{circ}}\equiv{\bar{l}}^2/(GM_{\rm{p}})$, where
\begin{equation}
\bar{l}=\eta\int l(t)\dot{M}_{\rm{trans,BH}}dt/M_{\rm{tot}}.
\end{equation}
After that, the matter losses angular momentum via viscosity and spread from $R_{\rm{circ}}$ to form an accretion disk. 

In table \ref{tab:1} we list different mass transfer scenarios under variance of disk parameters. We see from this table that the mass transfer scenarios are not sensitive to slight changes of disk orientation and inclination (see the definition of three CD configurations in the caption of table \ref{tab:1}). However, whether or not the mass transfer occurs depends strongly on the rotation direction of the CD: mass transfer is expected to occur when the CD is prograde, but not retrograde with the orbit. By study the spin-orbit coupling of this system, \cite{2014MNRAS.437.3255S} suggested the misalignment of the spin axis of LS 2883 and orbit axis is less than $90^\circ$, which supports a prograde CD.

\begin{table*}[]
\centering
\caption{\textbf{Different mass transfer scenarios under variance on the CD parameters.} \\
 $\bar{v}_{\rm{rel},\pm}$ and $T_{\pm}$ are the averaged relative velocity and the duration of mass transfer at the ``$\pm$" crossing respectively.
Configuration 1: $\theta_{\rm{n}}=40^\circ$, $\phi_{\rm{n}}=24^\circ$; Configuration 2: $\theta_{\rm{n}}=45^\circ$, $\phi_{\rm{n}}=19^\circ$; Configuration 3: $\theta_{\rm{n}}=35^\circ$, $\phi_{\rm{n}}=14^\circ$. Three values of each calculated parameters correspond to $n=3, n=3.3, n=3.5$. A bar in the value means mass transfer condition is not satisfied under the corresponding CD configuration. The bold values are used in the following parts of the paper. }\label{tab:1}
\label{my-label}
\begin{tabular}{|l|l|c|c|c|}
\hline
                            &                                             & \multicolumn{1}{l|}{Configuration 1} & \multicolumn{1}{l|}{\textbf{Configuration 2}} & \multicolumn{1}{l|}{Configuration 3} \\ \hline
\multirow{8}{*}{Prograde}   & $\bar{v}_{\rm{rel},+}$\,($\times10^7$ cm/s) & \textit{1.0}                     & \textbf{1.1}                     & \textit{1.2}                     \\ \cline{2-5}
                            & $\bar{v}_{\rm{rel},-}$                      & 1.3                     & \textbf{1.3}                     & 1.4                     \\ \cline{2-5}
                            & $T_+$\,(days)                               & 4.9, --, --                  & 7.7, --, --            & 8.7, --, --                 \\ \cline{2-5}
                            & $T_-$                                       & 8.6, 5.9, --                        & 9.5, \textbf{5.1}, --                      & 10.2, 3.5, --                        \\ \cline{2-5}
                            & $M_{\rm{tot},+}$\,($\eta\times10^{24}$\,g)      & 0.7, --, --                  & 0.8, --, --          & 1.2, --, --                  \\ \cline{2-5}
                            & $M_{\rm{tot},-}$                            & 3.4, 1, --                      & 2.8, \textbf{0.7}, --                      & 2.4, 0.4, --                      \\ \cline{2-5}
                            & $R_{\rm{circ},+}$\,($\times10^{10}$\,cm)    & 1.5, --, --                  & 1.1, --, --          & 0.85, --, --                    \\ \cline{2-5}
                            & $R_{\rm{circ},-}$                           & 1.8, 2.2, --                      & 1.2, \textbf{1.4}, --                      & 0.8, 1, --                      \\ \hline
\multirow{8}{*}{Retrograde} & $\bar{v}_{\rm{rel},+}$                      & 1.7                       & 1.8                              & 1.9                              \\ \cline{2-5}
                            & $\bar{v}_{\rm{rel},-}$                      & 2.7                     & 2.6                              & 2.4                              \\ \cline{2-5}
                            & $T_+$                                       & --                  & --                               & --                               \\ \cline{2-5}
                            & $T_-$                                       & 2.1, --, --                               & --                               & --                               \\ \cline{2-5}
                            & $M_{\rm{tot},+}$                            & --                  & --                               & --                               \\ \cline{2-5}
                            & $M_{\rm{tot},-}$                            & 0.1, --, --                               & --                               & --                               \\ \cline{2-5}
                            & $R_{\rm{circ},+}$                           & --                  & --                               & --                               \\ \cline{2-5}
                            & $R_{\rm{circ},-}$                           & 0.02, --, --                               & --                               & --                               \\ \hline
\end{tabular}
\end{table*}
For the purpose of further modeling, we work with $\theta_{\rm{n}}=45^\circ\,,\phi_{\rm{n}}=19^\circ$ and the assumption that the CD is prograde and $n=3.3$. In this case, $M_{\rm{tot}}=\eta7\times10^{23}$\,g and $R_{\rm{circ}}=1.4\times10^{10}$\,cm. $R_{\rm{circ}}$ is much larger than the radius of the light cylinder ($7.6\times10^7$\,cm), therefore the angular momentum of the accreted matter is enough to form an accretion disk. The mass transfer process lasts $\sim5$ days, while the flare lasts more than 60 days after the disk passage. Therefore the accretion rate of the disk should be much less than the mass transfer rate. Thus, as an approximation, we
consider the gravity-captured matter accumulated around the pulsar within $r_{\rm{BH}}$ during the mass transfer process. When the matter transfer ends, the accumulated matter begins to form an accretion disk.\\



\subsection{Evolution of the accretion disk}
As described above, after being captured by the gravity of the neutron star, complex physics processes (including shocks) will convert much of the kinetic energy of the gas into radiation and redistribute the remaining kinetic energy and angular momenta, before a torus of gas is formed at $R_{\rm{circ}}$. This phase is much alike the situation where a tidal disrupted star's debris is accreted onto a black hole \citep{1988Natur.333..523R,Cannizzo1990}. 

The torus is the predecessor of an accretion disk. After the accretion disk is developed, the inner edge decreases until it is disrupted by the torque of the magnetic field of the pulsar. The inner most radius $r_{\rm{M}}\approx\,r_{\rm{A}}$ \citep{2002apa..book.....F}, where
\begin{equation}
r_{\rm{A}}=5.1\times10^8\dot{M}_{\rm{acc,16}}^{-2/7}m_{\rm{p}}^{-1/7}\mu_{30}^{4/7}\,\text{cm}
\label{eqn:alfven}
\end{equation}
is the Alfven radius, $\dot{M}_{\rm{acc,16}}$ is the accretion rate of the accretion disk in units of $10^{16}$\,g/s, $\mu_{30}$ is the magnetic dipole in units of $10^{30}$\,G\,cm$^3$, $m_{\rm{p}}\equiv M_{\rm{p}}/M_\odot$. After the accretion disk reaches the $r_{\rm{M}}$, the propeller effect ejects the accreting matter thus the mass of the accretion disk declines steadily. This is much alike the situation where a pulsar is accreting from a fossil disk. Therefore, we follow the strategy of \cite{2000ApJ...534..373C}, to assume the accretion rate in two phases:
\begin{equation}
\dot{M}_{\rm{acc}}=\dot{M}_{\rm{acc,0}}\quad,\quad(r_{\rm{in}}>r_{\rm{M}})\nonumber
\end{equation}

\begin{equation}
\dot{M}_{\rm{acc}}=\dot{M}_{\rm{acc,0}}\big(\frac{t}{\tau}\big)^{-\beta}\quad,\quad(r_{\rm{in}}=r_{\rm{M}})
\label{eqn:accrate}
\end{equation}
where $t=0$ is the onset of disk formation and $t=\tau$ is the moment when the disk descends to $r_{\rm{M}}$, $\beta$ is the index to describe the decreasing of the accretion rate, $\beta=19/16$ for an electron scattering dominated disk opacity, and $\beta=1.25$ for a Kramer opacity. $\beta=5/3$ is a well-known value given by \cite{1988Natur.333..523R}, when considering a tidal disrupted star's debris accreting onto a black hole. $r_{\rm{in}}$ evolves with time as:
\begin{equation}
r_{\rm{in}}=R_{\rm{circ}}-v_rt\quad,\quad(t<\tau)\nonumber
\end{equation}
\begin{equation}
r_{\rm{in}}=r_{\rm{M}}\quad,\quad(t\ge\tau),
\label{eqn:rt}
\end{equation}

The constant accretion rate $\dot{M}_{\rm{acc,0}}$ is normalized to the total bounded mass:
\begin{equation}
M_{\rm{tot}}=\int^\infty_0\dot{M}_{\rm{acc}}dt,
\end{equation}
which gives:
\begin{equation}
\dot{M}_{\rm{acc,0}}=\frac{(\beta-1)M_{\rm{tot}}}{\beta\tau}.
\label{eqn:T1}
\end{equation}
On the other hand, if we assume the accretion disk to be a Shakura-Sunyaev disk \citep{1973A&A....24..337S},
then
\begin{eqnarray}
\tau&\approx&\frac{R_{\rm{circ}}}{v_r}\nonumber\\
&\approx&\frac{1}{2.7}\alpha^{-4/5}\dot{M}_{\rm{acc,16}}^{-3/10}m_{\rm{p}}^{1/4}R_{\rm{circ}}^{5/4}\times10^6\,\text{s}.
\label{eqn:T2}
\end{eqnarray}
where $\alpha$ is the viscosity index.

Combining equations (\ref{eqn:T1}, \ref{eqn:T2}), the accretion rate can be solved as:
\begin{equation}
\frac{\dot{M}_{\rm{acc,0}}}{10^{16}\,\rm{g}/\rm{s}}=\big[2.7\alpha^{4/5}\frac{\beta-1}{\beta}M_{\rm{tot,22}}m_{\rm{p}}^{-1/4}R_{\rm{circ,10}}^{-5/4}\big]^{10/7}
\label{eqn:mdotss}
\end{equation}
where $R_{\rm{circ,10}}$ is $R_{\rm{circ}}$ in units of $10^{10}$\,cm.

\begin{figure}
\centering
\includegraphics[width=8cm]{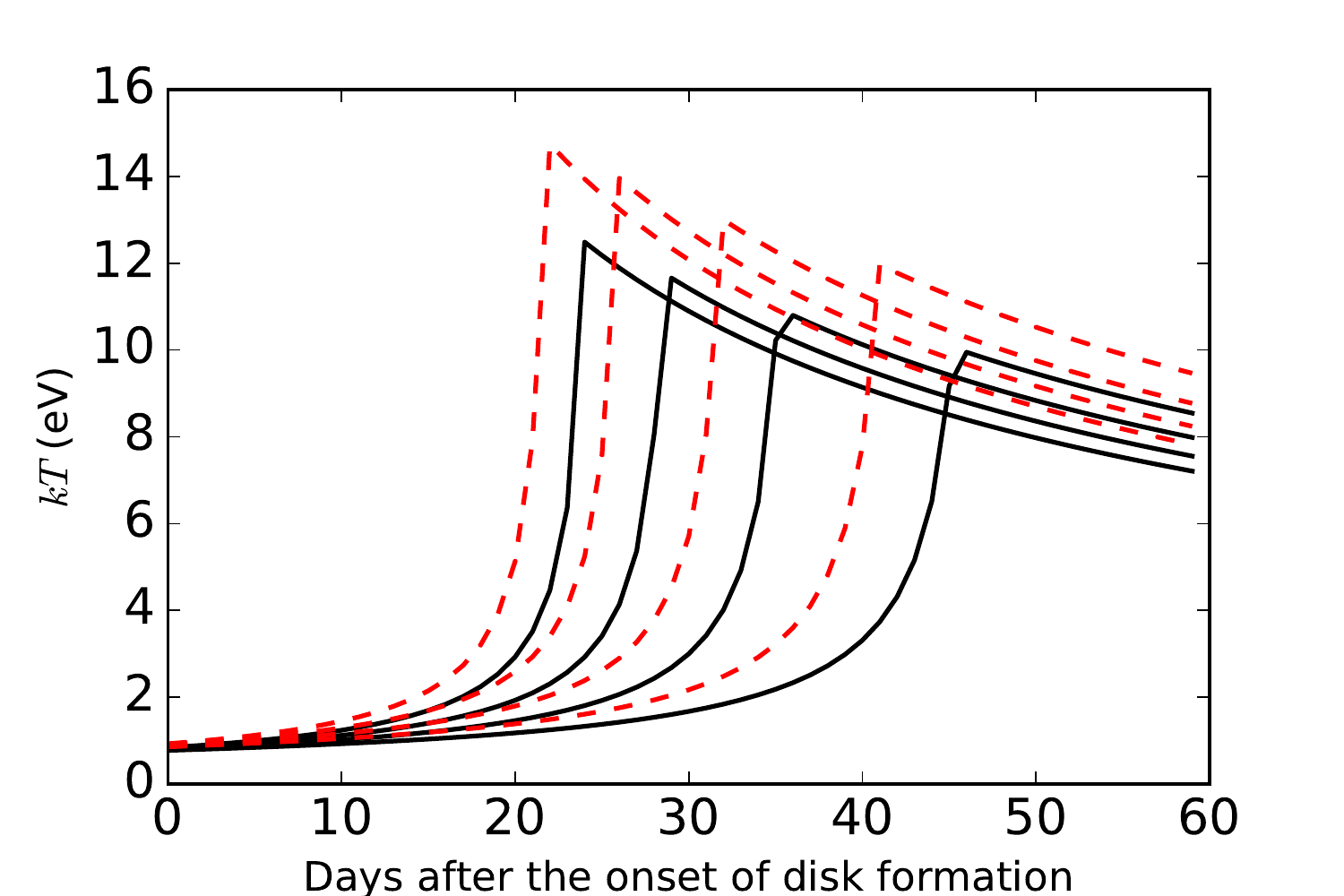}
\caption{\textbf{The temperature of the inner most region of the accretion disk, as a function of the time after the formation of the accretion disk:} From left to right correspond to $\alpha=0.35, 0.3, 0.25, 0.2$. The solid lines are for $\beta=19/16$, and the dashed lines are for $\beta=1.25$. For all curves, $\eta=0.05$.}
\label{fig:2}
\end{figure}

\begin{figure}
\centering
\includegraphics[width=8cm]{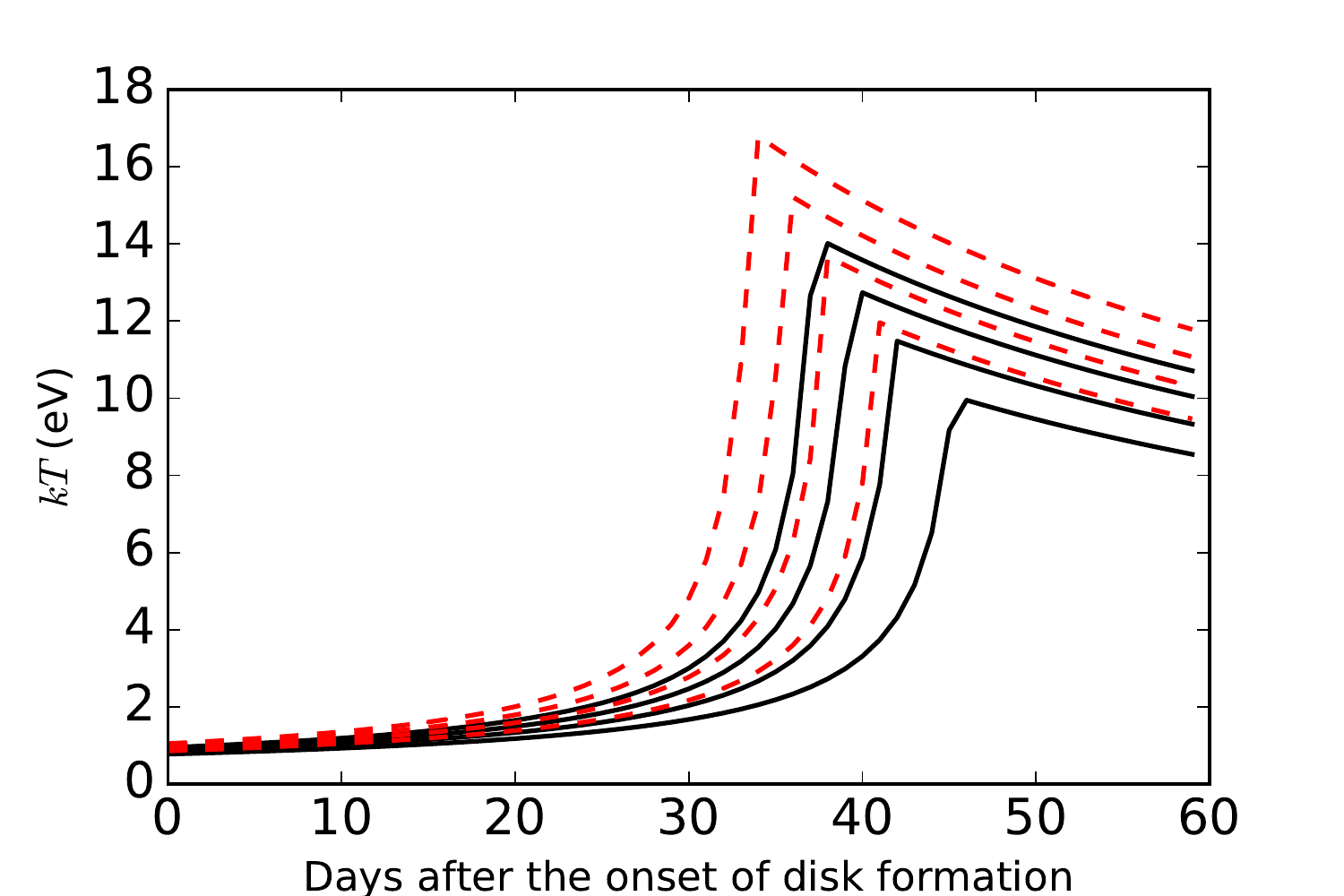}
\caption{\textbf{The temperature of the inner most region of the accretion disk, as a function of the time after the formation of the accretion disk:} From left to right correspond to $\eta=0.08, 0.07, 0.06, 0.05$. The solid lines are for $\beta=19/16$, and the dashed lines are for $\beta=1.25$. For all curves $\alpha=0.2$.}
\label{fig:3}
\end{figure}

Since the temperature of the accretion disk increase inwardly, the inner most region of the accretion disk dominates the radiation, where the temperature is:
\begin{equation}
T=1.4\times10^4\alpha^{-1/5}\dot{M}_{\rm{acc,16}}^{3/10}m_{\rm{p}}^{1/4}r_{\rm{in,10}}^{-3/4}\,\text{K},
\label{eqn:temper}
\end{equation}
where $r_{\rm{in,10}}$ is the inner radius of the accretion disk in units of $10^{10}$\,cm.
With the time evolution of $\dot{M}_{\rm{acc}}$ and $r_{\rm{in}}$ from equations (\ref{eqn:T1}, \ref{eqn:rt}), the temperature as a function of $t$ is shown in figures \ref{fig:2} and \ref{fig:3}.

\subsection{Inverse Compton scattering from the pulsar wind}
\begin{figure}
\centering
\includegraphics[width=8cm]{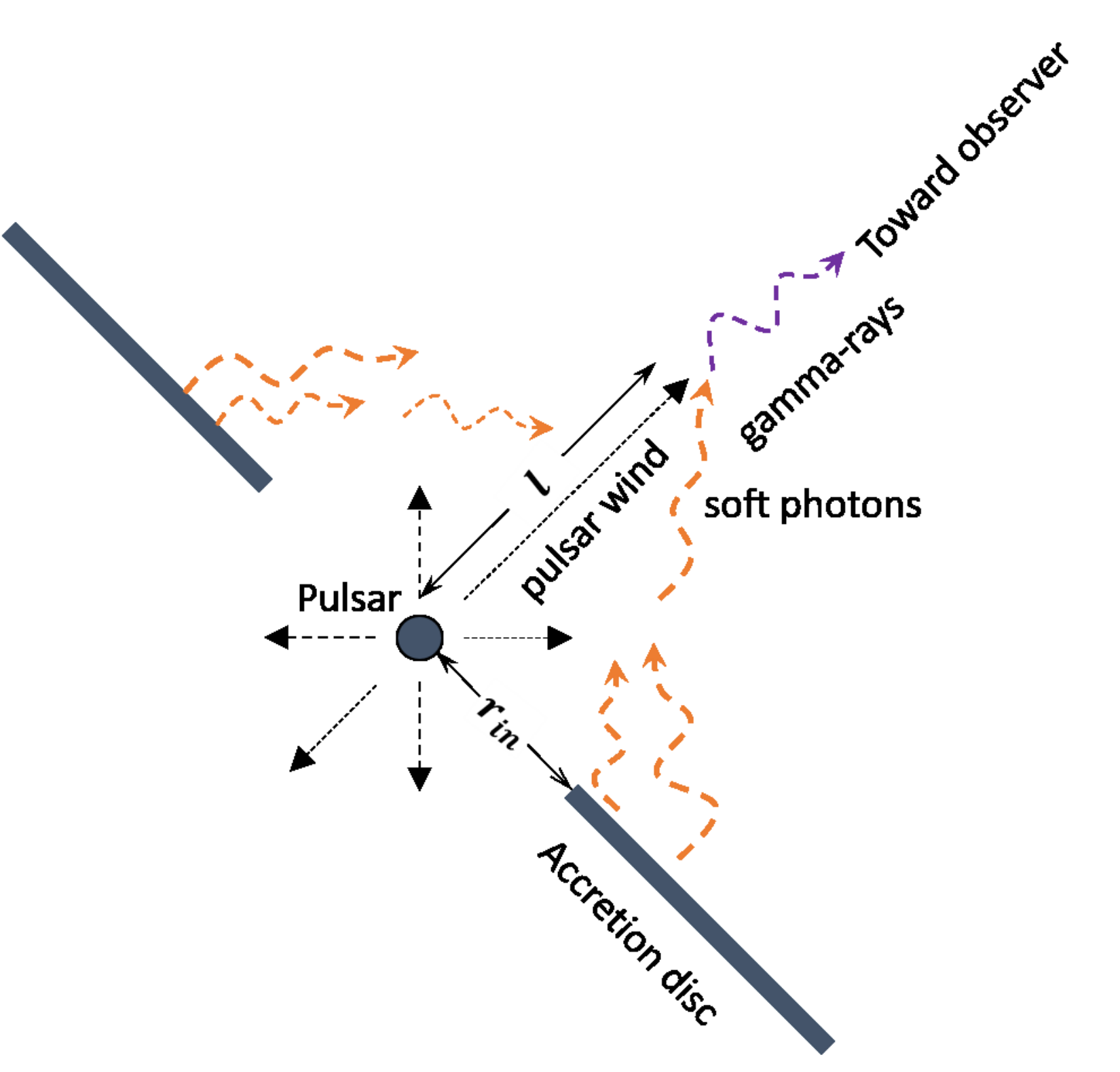}
\caption{\textbf{Illustration of the IC process:} Soft photons are emitted from the inner region of the accretion disk around the pulsar, the relativistic electrons in the pulsar wind up-scatter the soft photons to gamma-rays. The inclination of the accretion disk is not necessarily face-on as shown in this figure.}\label{fig:4}
\end{figure}

The soft photons from the accretion disk are inverse-Compton scattered to high energy by electrons in the pulsar wind. If we assume a monochromatic electron energy of the pulsar wind, with the Lorentz factor $\Gamma$, then the characteristic energy of the scattered photon is \citep{1970RvMP...42..237B}:\\
\begin{equation}
E_{\gamma}\approx300(kT/1\,\text{eV})\Gamma_{0,4}^2\,\text{MeV},
\label{eqn:E1}
\end{equation}
in the Thomson regime ($kT\Gamma/0.511\,\text{MeV}\ll1$),
and
\begin{equation}
E_{\gamma}\approx5.11\Gamma_{0,4}\,\text{GeV},
\label{eqn:E2}
\end{equation}
in the Klein-Nishina regime ($kT\Gamma_0/0.511\,\text{MeV}\gtrsim1$),
where $\Gamma_4\equiv\Gamma/10^4$. For $kT\sim10$\,eV as we calculated above, a pulsar wind with $\Gamma\sim10^3-10^4$ will produce inverse Compton scattered photons in the {\it{Fermi}}-LAT's energy range (100\,MeV--100\,GeV). Assuming an isotropic soft photons field, the energy differential power distribution (photons per energy range, per unit time) of inverse Compton scattering in Thomson limit is \citep{1970RvMP...42..237B}:
\begin{equation}
\dot{N}_{\gamma}=\int dN_{\rm{e}}(l)\int^\infty_0\frac{2}{\Gamma^2\epsilon_{\rm{ph}}}\pi r_0^2cn(\epsilon_{\rm{ph}},l)f(x)d\epsilon_{\rm{ph}},
\label{eqn:power}
\end{equation}
where $dN_{\rm{e}}(l)$ is the number of electrons at distance $l$ along the line of sight, within range $dl$. In figure \ref{fig:4}, we illustrate the IC process. The accretion disk is not necessarily face on as plotted; If we suppose nearly all the spin down energy of the pulsar is converted into the kinetic energy of electrons in the pulsar wind at $l$ (as adopted by \cite{1997ApJ...477..439T}, $\sim96\%$ of the pulsar wind energy is in the kinetic energy of electrons for this system. For a discussion on the role of pulsar wind in spin braking, see \citealt{2016SCPMA..59a5752T} and references therein), then we have:
\begin{equation}
dN_{\rm{e}}(l)=\frac{L_{\rm{spin}}}{\Gamma m_{\rm{e}}c^2}\frac{dl}{c}.
\label{eqn:Ne}
\end{equation}
$\epsilon_{\rm{ph}}$ in equation (\ref{eqn:power}) is the energy of the soft photons, $r_0$ is the classic radius of the electron, $n(\epsilon_{\rm{ph}},l)$ is energy differential number density of the soft photons at $l$, $f(x)=2x\ln x+x+1-2x^2$ for $0<x<1$, $f(x)=0$ for $x>1$ and $x\equiv E_{\gamma}/(4\Gamma^2\epsilon_{\rm{ph}})$.

For simplicity, we consider the case where the accretion disk is viewed face-on. The soft photon field at $l$ is:
\begin{equation}
n(\epsilon_{\rm{ph}},l)=\frac{4\pi l}{h^3c^3}\int^{R_{\rm{out}}}_{R_{\rm{in}}}\frac{\epsilon_{\rm{ph}}^2RdR}{(R^2+l^2)^{3/2}(\exp\frac{\epsilon_{\rm{ph}}}{kT(R)}-1)}.
\label{eqn:ng}
\end{equation}
A change of the inclination will slightly reduce the density of photon field. This will eventually lead to larger fitted $\eta$.
Combining equations (\ref{eqn:power},\ref{eqn:Ne},\ref{eqn:ng}), and work out the integrate over $l$,
\begin{equation}
\dot{N}_{\gamma}=\frac{8\pi^2r_0^2}{h^3c^3}\frac{L_{\rm{spin}}}{\Gamma^3m_{\rm{e}}c^2}\int \epsilon_{\rm{ph}}f(x)\int_{R_{\rm{in}}}^{R_{\rm{out}}}\frac{dR}{\exp\frac{\epsilon_{\rm{ph}}}{kT(R)}-1}d\epsilon_{\rm{ph}},
\label{eqn:33}
\end{equation}

Since the surface brightness of a black body $\propto T^4$, and $T\propto R^{-3/4}$, the soft photon field from the accretion disk can be represented by a ring with width of $r_{\rm{in}}$ and temperature of $T(r_{\rm{in}})$. Therefore the integration over $R$ in Equation (\ref{eqn:33}) can be simplified as:
\begin{equation}
\dot{N}_{\gamma}=\frac{8\pi^2r_0^2}{h^3c^3}\frac{L_{\rm{spin}}r_{\rm{in}}}{\Gamma^3m_{\rm{e}}c^2}\int\frac{\epsilon_{\rm{ph}}f(x)}{\exp\frac{\epsilon_{\rm{ph}}}{kT(r_{\rm{in}})}-1}d\epsilon_{\rm{ph}},
\label{eqn:34}
\end{equation}

Assuming the pulsar wind is isotropic, the spectrum of the inverse Compton process is:
\begin{equation}
F_{\gamma}=\frac{\dot{N}_{\rm{\gamma}}}{4\pi D^2},
\label{eqn:f}
\end{equation}

\begin{figure}
\centering
\includegraphics[width=8cm]{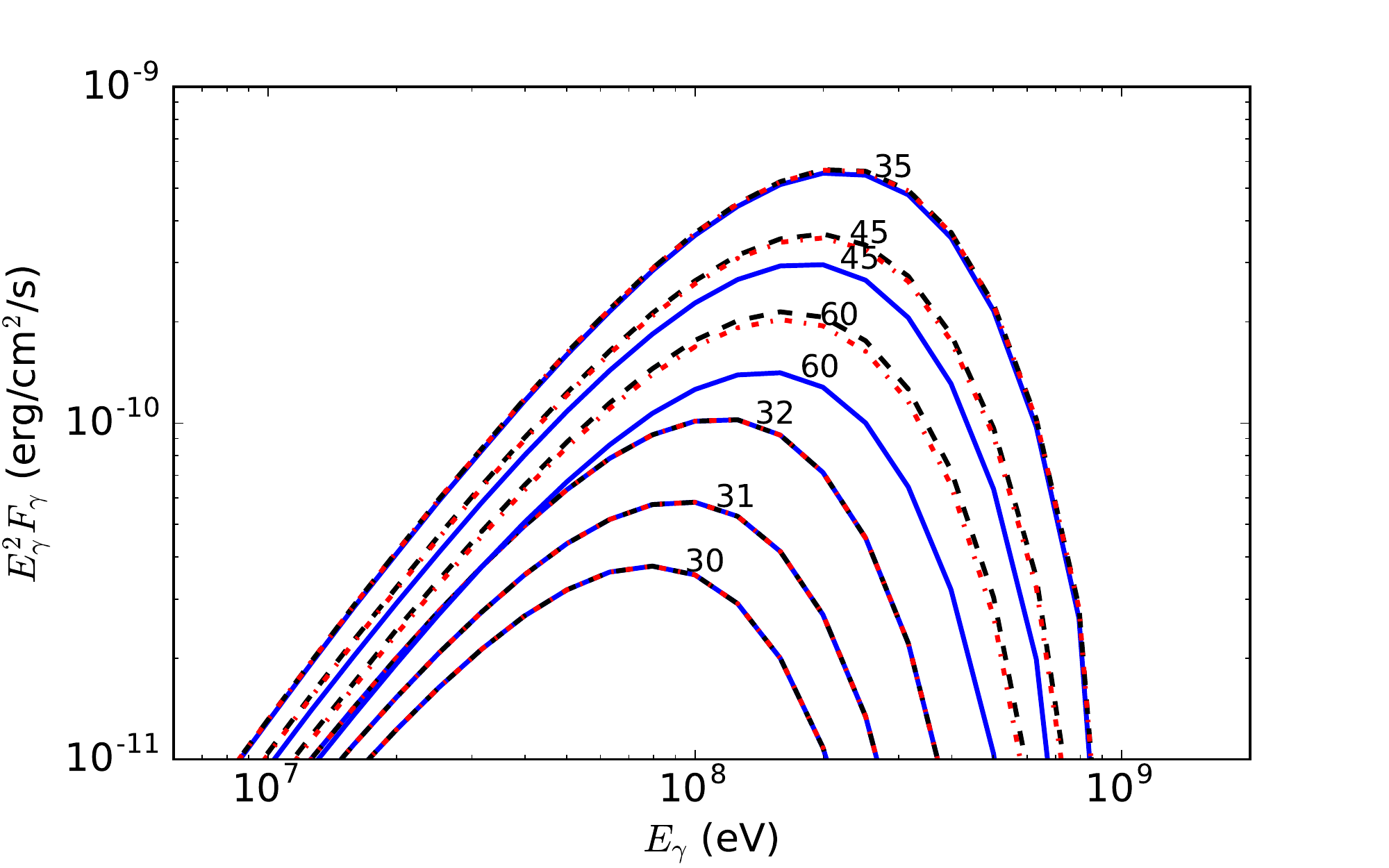}
\caption{\textbf{The evolution of the SED of IC:} black and dashed curves correspond to $(\beta, \eta)=(19/16, 0.063)$, red and dash-dotted curves correspond to $(\beta, \eta)=(1.25, 0.05)$ and blue and solid curves correspond to $(\beta, \eta)=(5/3, 0.025)$; $\alpha=0.2$ and $\Gamma=1.5\times10^3$ for all curves. The numbers labeled besides the apexes of each group of curves indicates the corresponding days after the periastron. On the day 30, 31, 32 and 35, the SEDs of three colors superimpose on each other. See text for detail of this plot.}
\label{fig:SED}
\end{figure}

With above equations, and the known temporal function of $T$ and $r_{\rm{in}}$ in equations (\ref{eqn:temper}) and (\ref{eqn:rt}), we know how the spectrum energy distribution (SED) evolves with time, as shown in figure \ref{fig:SED}. For each time, we show the SED corresponding for $(\beta, \eta)=(19/16, 0.059), (1.25, 0.047), (5/3, 0.024)$ in black dashed, red dash-dotted and blue solid curves respectively. $\alpha$ is 0.4 as the best fitted value found in the next section. The numbers labeled besides the apexes of each group of curves indicates the corresponding days after the periastron.

\section{Comparing with observations}
\begin{figure*}
\centering
\includegraphics[width=16 cm]{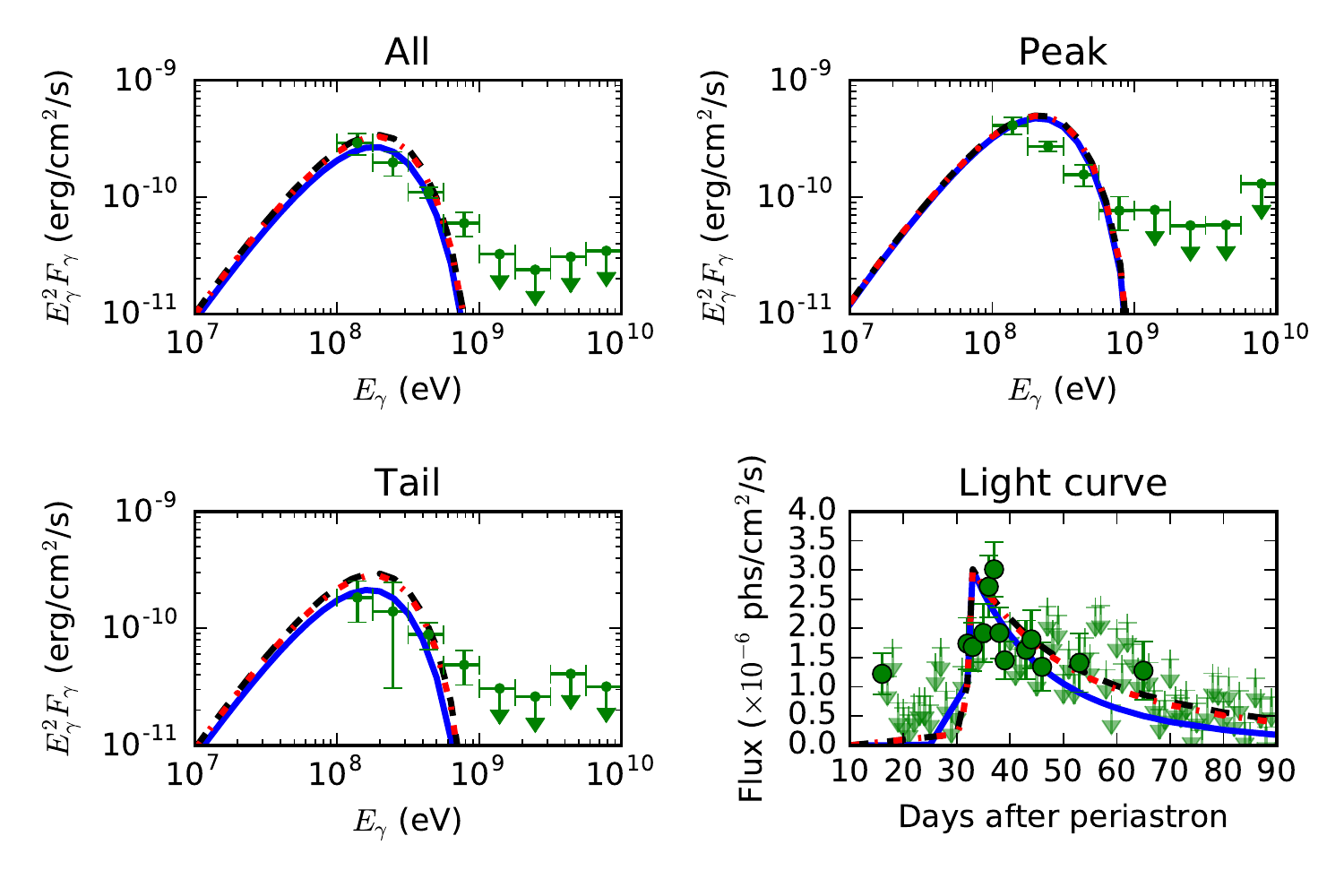}
\caption{\textbf{SEDs and light curve of the gamma-ray flare at 2010/2011}: The panels titled with ``All" ``Peak" and ``Tail" corresponds to SEDs averaged over the whole durations, the peak and the tail phases. The black dashed, red dash-dotted and blue solid curves are SEDs calculated with $(\beta, \eta)=(19/16, 0.067), (1.25, 0.053), (5/3,0.026)$; $\alpha$ is 0.2 and $\Gamma=1.5\times10^3$ as the best fitted value. The green points with error bars are observed SED data from \protect\cite{2015ApJ...811...68C}, those points with downward arrows are observational upper limits. The panel titled with ``Light curve" are calculated light curves and the observed one. The markers and the line-styles of the light curves are the same as in other panels.}
\label{fig:ffo}
\end{figure*}
\begin{figure*}
\centering
\includegraphics[width=16 cm]{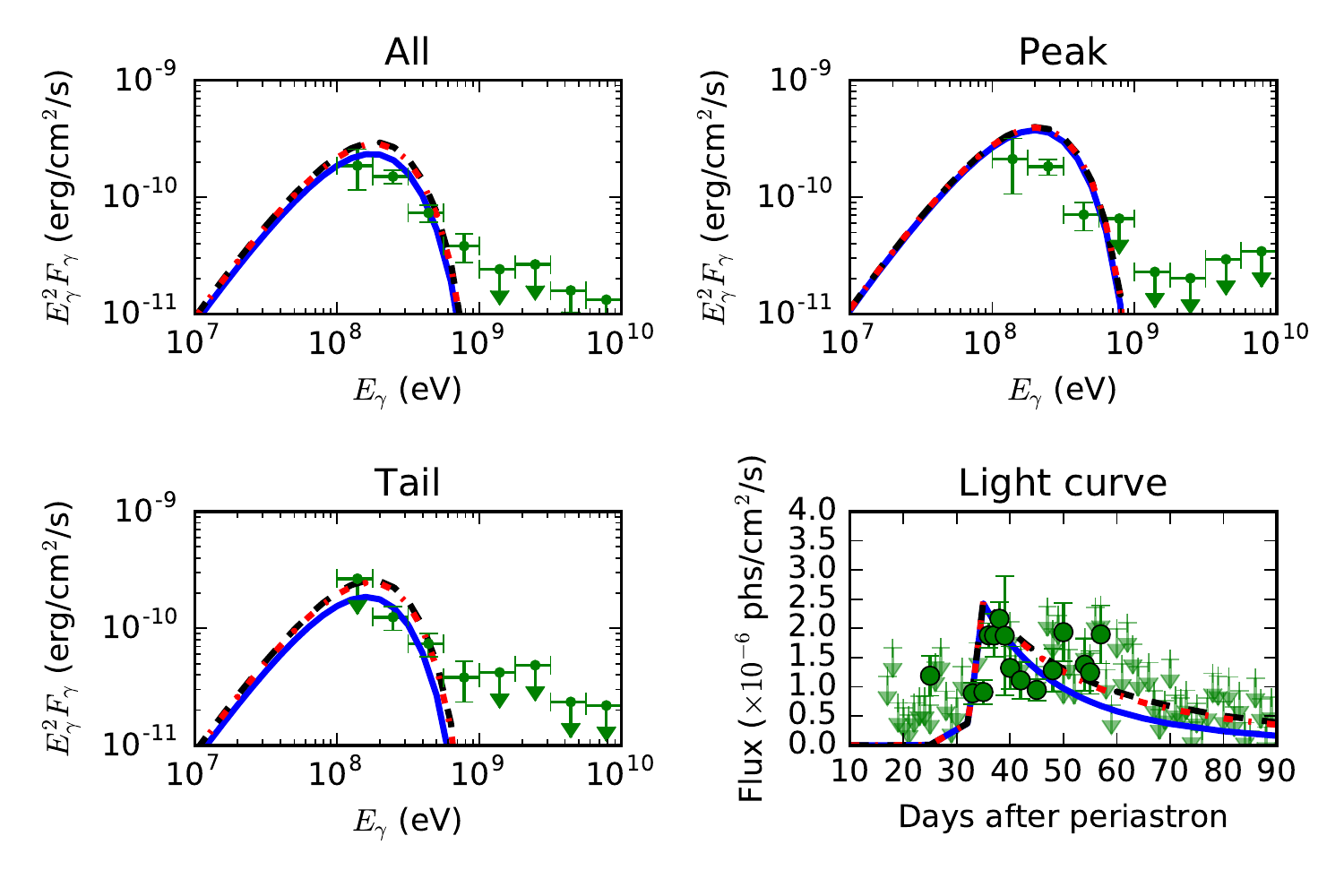}
\caption{\textbf{SEDs and light curve of the gamma-ray flare at 2014}: The panels titled with ``all" ``Peak" and ``Tail" corresponds to SEDs averaged over the whole durations, peak and tail phases. The black dashed, red dash-dotted, blue solid curves are SEDs calculated with $(\beta, \eta)=(19/16, 0.063), (1.25, 0.05), (5/3,0.025)$; $\alpha$ is 0.2 and $\Gamma=1.5\times10^3$ as the best fitted value. The green points with error bars are observed SED data from \protect\cite{2015ApJ...811...68C}, those points with downward arrows are observational upper limits. The panel titled with ``Light curve" are calculated light curves and the observed one. The markers and the line-styles of the light curves are the same as in other panels.}
\label{fig:fft}
\end{figure*}

In order to evaluate this model, we compare it against observations. \cite{2015ApJ...811...68C} presented the light curves and SEDs of the gamma-ray flares in 2010/2011 and 2014 passages. They divided the light curves of the flares into two phases based on their fluxes level: the peak phase and the tail phase. The peak phase of the first/second passage is from the 31\,st day to 40\,th/42\,nd day after the periastron, and the tail phase is from the 40\,th/42\,nd to 71st day after the periastron. They obtained the SEDs by integrating the photons over the peak phases, tail phases and all the flares duration. To compare with them, we also average our calculated SEDs over the corresponding time ranges. In figures \ref{fig:ffo} and \ref{fig:fft}, sub-panels titled with ``All", ``Peak" and ``Tail" correspond to SEDs averaged over the whole durations, peak and tail phases. The black dashed, red dash-dotted and blue solid curves are SEDs calculated with different pairs of $(\beta, \eta)$ (see the captions of the figures for details). The green points with error bars are observed SED data from \cite{2015ApJ...811...68C}, those points with downward arrows are observational upper limits. Integrating equation (\ref{eqn:f}) over the {\it Fermi}-LAT energy range (100\,MeV-100\,GeV) at each instance gives the flux as function with time, i.e., the light curves. We compare the calculated light curves with the observed ones in the panels titled with ``Light curve" of figures \ref{fig:ffo} and \ref{fig:fft}.

\section{conclusion and discussion}
In this paper, we study a new explanation of the observed post-periastron GeV flares of the binary system PSR B1259-63/LS2883. This phenomenon could be results as the pulsar wind inverse Compton-scattering of the soft photons, which are emitted from an accretion disk around the pulsar. The accretion disk composed by the matter from the circumstellar disk (CD) captured by the pulsar's gravity at disk-crossing. The pulsar crosses the CD twice in each orbital period, one pre-periastron and another post-periastron. In the post-periastron crossing, the density of the CD is not enough so that the pulsar wind prevents the matter from being accreted. While in the other crossing the density is sufficient. That explain the GeV flares are only observed once in each orbit.

With certain parameters of the accretion disk, this model can reproduce the observed SEDs and light curves satisfactorily (see figures \ref{fig:ffo} and \ref{fig:fft}).
\subsection{Sub-structures in the light curves and the disk instability}
It is worth mentioning that the light curves in \cite{2015ApJ...811...68C} show some significant sub-structures other than a pure decay. The disk instability might produce the above-mentioned small-scaled outbursts (see \citealt{1989ApJ...343..241M} and references therein; see \citealt{2001NewAR..45..449L} for a review). The recurrence time of those outbursts is of the viscous timescale near the inner edge of the disc, which is $\sim0.1$\,days estimated with the best fitted parameters. However the variance time scale of the sub-structures in the light curves is of several days. Therefore, if the disk instability accounts for the sub-structures in the light curves, a different $\alpha\lesssim0.05$ is needed in the decaying phase than that in the rising phase. It is possible that the $\alpha$ of an accretion disk is not a constant and is time-dependent via other disk properties \citep{1984AcA....34..161S}. In this work, we only consider the quasi-stable disk as a simplification and do not try to reproduce the sub-structures.
\subsection{UV excess from the accretion disk}
One critical evidence to prove or to disprove the formation of the accretion disk is the expected UV emission excess.
The frequency differential energy flux at frequency $\nu$ from the accretion disk is:
\begin{equation}
F_{\nu}=\frac{4\pi h\nu^3\cos i}{c^2D^2}\int^{R_{\rm{out}}}_{R_{\rm{in}}}\frac{RdR}{\exp h\nu/kT -1},
\end{equation}
where $i$ is the inclination of the disk and $D$ is the distance to the system. We assume $i=45^\circ$ and use the simplification that the integration over the whole accretion disk can be approximated by a isothermal ring with width $r_{\rm{in}}$ and temperature $T(r_{\rm{in}})$.
While the flux from the Be star is:
\begin{equation}
F_{\nu,\star}=\frac{2h\nu^3\pi R^2_\star}{c^2D^2}\frac{1}{\exp h\nu/kT_\star-1},
\end{equation}
where $T_\star=30200$\,K is the temperature of the Be star,
and the flux from the CD $F_{\nu,\rm{sd}}$ can be represented by an isothermal disk with a temperature of $0.6\,T_\star$ and an effective radius of the pseudophotosphere $R_{\rm{eff}}(\nu)$ \citep{Carciofi2006}.

\begin{figure}
\centering
\includegraphics[width=8cm]{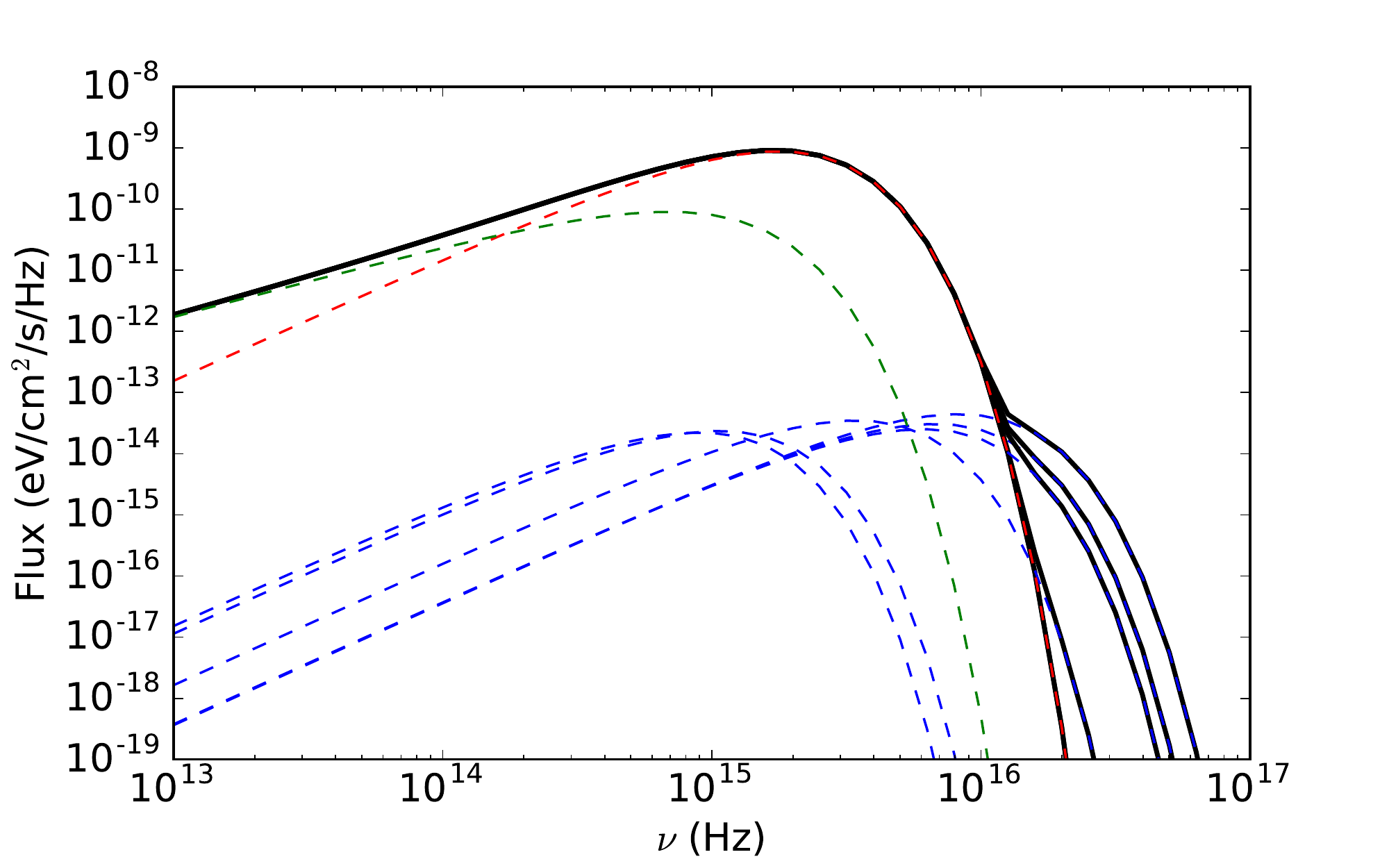}
\caption{\textbf{IR/Optical/UV SED from the system:} The solid curves are overall SED, the dashed curves are from individual components. Green: from the circumstellar disk; Red: from the Be star; blue: emission from accretion disk from 10 to 60 days after periastron, with parameters $(\beta, \sigma, \alpha)=(1.25, 0.05, 0.2)$. }\label{fig:5}
\end{figure}

The SED in the optical band is shown in figure \ref{fig:5}. An excess of the SED in UV band is a natural prediction. The energy flux ($\nu F_{\nu}$) from $2\times10^{16}$\,Hz (82.7 eV) to $3\times10^{16}$\,Hz (165 eV) is from $3.2\times10^{-11}$\,ergs/cm$^2$/s to $\sim1\times10^{-12}$\,ergs/cm$^2$/s at the peak of the flare. It is higher than the low-energy tail of the soft X-ray emission from the shock, extrapolating to the same energy range from the figure 3 of \cite{2012ApJ...753..127K}.
Therefore, an extreme-UV observation covering $2\times10^{16}-3\times10^{17}$\,Hz (10-15 nm) during the GeV flare period can be used to test the proposed model.
\subsection{The dependence on the CD density profile of the model}
We see from table \ref{tab:1} that the mass transfer scenario dependents on the density profile of the CD. If $n=3$, or equivalently $\rho_0$ halves from the current assumed value of $1\times10^9$\,g/cm$^3$, the mass transfer condition will not be satisfied at disk-crossing; on the other hand, if $n=3.5$ or equivalently $\rho_0$ doubles from the current value, the mass transfer can occur at both disk-crossings. Thus people expect a smaller GeV peak $\sim20$\,days after the main flare with $\sim1/4$ of the flux. The fluxes uncertainties of the current light curves are too large to confirm or to exclude it.
\subsection{The accretion of angular momentum}
In our treatment, the accretion of angular momentum is due to the velocity and density gradient along the CD. Early studies suggested the specific angular momentum captured in this way is equation (\ref{eqn:specl}). However \cite{1980MNRAS.191..599D} and \cite{1986MNRAS.218..593L} argued analytically and numerically respectively that, very little (if not none) angular momentum can be accreted by the center mass in this way. Later more sophisticated numerical calculations \citep{1995A&A...295..108R,1997A&A...317..793R} showed that up to 70\% of the total available angular momentum can be accreted. In our paper, as long as the actual specific angular momentum is enough, so that the circular radius $R_{\rm{circ}}$ is well outside the light cylinder, the validity of our model is not damaged. In our model, $R_{\rm{circ}}$ influences the gamma-ray flux in the combination $\alpha^{4/5}R^{-5/4}_{\rm{circ}}$, thus any change of $R^{-5/4}_{\rm{circ}}$ will be absorbed by a change of $\alpha$ during the fitting.

In fact, as $\alpha$ varies in a reasonable range, the time delay between the disk-crossing and the flare can be adjusted accordingly in a range of a couple of weeks. As a result, this model has a weak predictive power over the position of the CD. As we mentioned in above paragraphs, that this model is potentially testable with observation of the UV excess. Besides, we expect that the spin period derivative of PSR B1259-63 increases during the GeV flare, due to the additional torque acting on the pulsar magnetic fields by the accretion disk.
\acknowledgments
SXY thanks Prof. J. Takata for helpful instructions. The authors thanks anonymous referee for his/her careful reviewing and helpful comments, which brings much improvement of the manuscript.
This work is partially supported by a GRF grant under 17302315.

\end{document}